\documentclass[preprint]{aastex}

\newcommand{\cxo}{{\sl Chandra}}
\newcommand{\etal}{et~al.}
\newcommand{\ltapprox}{\hbox{\raise0.5ex \hbox{$<$}
  \kern-1.1em \lower0.5ex \hbox{$\sim$}}}
\newcommand{\snr}{G266.2$-$1.2}
\newcommand{\twomass}{{\sl 2MASS}}

\slugcomment{ Accepted for publication in the Astrophysical Journal }

\shorttitle{On the Expansion Rate, Age, and Distance of \snr}
\shortauthors{Allen \etal}

\begin{document}


\title{On the Expansion Rate, Age, and Distance of the Supernova Remnant
  \snr\ (Vela~Jr.)}

\author{G. E. Allen}
\affil{MIT Kavli Institute for Astrophysics and Space Research, 77
  Massachusetts Avenue, NE83-557, Cambridge, MA 02139}
\email{gea@space.mit.edu}

\author{K. Chow}
\affil{Weston High School, 444 Wellesley Street, Weston, MA 02493}
\email{kc71135@gmail.com}

\author{T. DeLaney}
\affil{Department of Physics and Engineering, West Virginia Wesleyan
  College, Box 112, 59 College Avenue, Buckhannon, WV 26201}
\email{delaney\_t@wvwc.edu}

\author{M. D. Filipovi\'{c}}
\affil{University of Western Sydney, Locked Bag 1797, Penrith South DC,
  NSW 1797, Australia}
\email{m.filipovic@uws.edu.au}

\author{J. C. Houck}
\affil{MIT Kavli Institute for Astrophysics and Space Research, 77
  Massachusetts Avenue, NE80-6023, Cambridge, MA 02139}
\email{houck@space.mit.edu}

\author{T. G. Pannuti}
\affil{Space Science Center, Department of Earth and Space Sciences,
  Morehead State University, 235 Martindale Drive, Morehead, KY 40351}
\email{t.pannuti@moreheadstate.edu}

\and

\author{M. D. Stage}
\affil{University of Massachusetts, Department of Astronomy, LGRT-B 619E,
  710 North Pleasant Street, Amherst, MA 01003-9305}
\email{mikstage@astro.umass.edu}


\begin{abstract}

An analysis of \cxo\ ACIS data for two relatively bright and narrow portions
of the northwestern rim of \snr\ (a.k.a.\ RX~J0852.0-4622 or Vela~Jr.)
reveal evidence of a radial displacement of $2.40 \pm 0.56$~arcsec between
2003 and 2008.  The corresponding expansion rate ($0.42 \pm
0.10$~arcsec~yr$^{-1}$ or $13.6 \pm 4.2$\%~kyr$^{-1}$) is about half the
rate reported for an analysis of {\sl XMM-Newton} data from a similar, but
not identical, portion of the rim over a similar, but not identical, time
interval \citep[$0.84 \pm 0.23$~arcsec~yr$^{-1}$,][]{kat08a}.  If the \cxo\
rate is representative of the remnant as a whole, then the results of a
hydrodynamic analysis suggest that \snr\ is between 2.4 and 5.1~kyr old if
it is expanding into a uniform ambient medium (whether or not it was
produced by a Type Ia or Type II event). If the remnant is expanding into
the material shed by a steady stellar wind, then the age could be as much as
50\% higher.  The \cxo\ expansion rate and a requirement that the shock
speed be greater than or equal to 1000~km~s$^{-1}$ yields a lower limit on
the distance of 0.5~kpc. An analysis of previously-published distance
estimates and constraints suggests \snr\ is no further than 1.0~kpc. This
range of distances is consistent with the distance to the nearer of two
groups of material in the Vela Molecular Ridge \citep[$0.7 \pm
0.2$~kpc,][]{lis92a} and to the Vel~OB1 association
\citep[0.8~kpc,][]{egg82a}.

\end{abstract}




\section{Introduction}

The shell-type supernova remnant \snr\ was discovered in the {\sl ROSAT}
all-sky survey data and, based upon its equatorial coordinates, named
RX~J0852.0$-$4622 \citep{asc98a}.  It lies along the same line of sight as
the Vela supernova remnant, which is considerably brighter than \snr\
(Vela-Z) at radio frequencies \citep{mil68a,boc98a,com99a} and at X-ray
energies below 1~keV \citep{asc98a}.  For these reasons, it is not
surprising that \snr\ (aka ``Vela Jr.'') was only recently identified as a
separate object.%
\footnote{\cite{wan02a} consider the possibility that \snr\ may not be a
  separate object, but is instead part of the Vela supernova remnant,
  produced by a fast-moving ejecta clump that has interacted with the shell.
  Here, we assume that \snr\ and Vela are separate objects because the
  spectral properties of \snr\ and the Vela ejecta clumps ``A'' and ``D''
  are not the same. Features ``A'' \citep{miy01a,kat06a} and ``D''
  \citep{plu02a,kat05b} exhibit thermal X-ray spectra, while X-ray spectra
  of \snr\ seem to be entirely nonthermal.  Furthermore, \snr\ is a source
  of TeV gamma rays, and at least feature ``D'' is not \citep[][Figs.~1, 3,
  4]{aha07a}.}

The remnant is nearly circular with a relatively large angular radius of
about 0.9$^{\circ}$. It has relatively low radio \citep{dun00a} and X-ray
surface brightnesses. To the extent that it is possible to distinguish the
emission of \snr\ from the emission of Vela, the X-ray
\citep{sla01a,bam05a,pan10a} and radio spectra of \snr\ appear to be
dominated by synchrotron radiation \citep[or, perhaps, jitter
radiation,][]{oga07a} from TeV and GeV electrons, respectively.  Images in
these two wavelengths are fairly similar to one another \citep{stu05a} and
to a TeV gamma-ray image \citep{aha07a}.  The detection of TeV gamma rays
\citep{kat05a,aha05a,eno06a,aha07a} provides unequivocal evidence of the
presence of TeV cosmic rays.  \cite{tan11a} and \cite{lan12a} report the
detection of \snr\ in GeV gamma rays. However, images at this energy differ
somewhat from the TeV gamma-ray, X-ray, and radio images, perhaps due to a
spatially unresolved (at GeV energies) combination of emission from \snr\
and from the pulsar wind nebula of PSR~J0855$-$4644.

The 64.7~ms radio pulsar PSR~J0855$-$4644 \citep{kra03a}, which lies near
the southeastern rim of \snr, is associated with an X-ray--emitting
\citep{ace13a} and, perhaps, gamma-ray--emitting
\citep{aha07a,tan11a,lan12a} pulsar wind nebula. However, \cite{ace13a}
argue that the idea that PSR~J0855$-$4644 and \snr\ are associated with each
other \citep{red05a} is unlikely because the space velocity of the pulsar
would have to be unusually high and because there is no evidence of a bow
shock or a trail around it.  Furthermore its characteristic age
\citep[140~kyr,][]{kra03a} exceeds our age range for \snr\ (2.4--5.1~kyr) by
more than an order of magnitude.

There is a compact X-ray--emitting object (CXOU~J085201.4--461753) near the
geometric center of \snr\ \citep{asc98a,sla01a,mer01a,pav01a,kar02a} at
least in the plane of the sky. This compact object and the supernova remnant
may also lie at the same distance since (model-dependent) estimates of the
absorption column density to the object \citep[$n_{\rm H} \sim 0.3$--$1.1
\times 10^{22}\ {\rm cm}^{-2}$,][]{pav01a,kar02a,bec06a} are similar to
estimates of the absorption column density to the remnant \citep[$n_{\rm H}
\sim 0.1$--$1.1 \times 10^{22}\ {\rm
cm}^{-2}$,][]{asc98a,tsu00a,sla01a,iyu05a,bam05a, ace13a}.%
\footnote{While the association of CXOU~J085201.4--461753 and \snr\ seems
  plausible, we cannot exclude the possibility that they are unrelated
  because the estimates of the absorption column densities are uncertain and
  because this region of the sky is fairly busy.}
The lack of evidence of R band emission from CXOU~J085201.4--461753, down to
a limiting magnitude of about 25.6 \citep{mig07a}, suggests that it was
produced by the collapse of a massive star \citep{pav01a}. There is no
evidence of X-ray pulsations
\citep{bec06a} or of a known radio pulsar at this location.%
\footnote{See the ATNF Pulsar Catalogue at
  http://www.atnf.csiro.au/people/pulsar/psrcat/ \citep{man05a}.}
The object does lie in or near a 6~arcsec-diameter optically-emitting nebula
\citep{pel02a,mig07a}, at least on the plane of the sky, but it is not clear
that the detected optical (H$\alpha$ or [N{\sc{II}}]) source and the X-ray
source are the same object \citep{mig09a}.

The complexity of the region makes it difficult to make a strong statement,
but \snr\ does not appear to be a source of neutral hydrogen
\citep{dub98a,tes06a}, infrared \citep{nic04a}, [O{\sc{III}}]
\citep{fil01a}, or far-ultraviolet \citep{nis06a,kim12a} emission.

We used the \cxo\ telescope to observe the thin filaments in the bright
northwestern region of \snr\ on two separate occasions.  These data and the
techniques used to analyze them are described in \S\ref{data}. Constraints
on the age and distance of the remnant are discussed in \S\ref{dis}. The
conclusions are presented in \S\ref{concl}.


\section{Data and Analyses}
\label{data}

The northwestern region of \snr\ was observed with \cxo%
\footnote{The \cxo\ {\sl X-ray Observatory} is described in the \cxo\
  Proposers' Observatory Guide, which is available at
  http://cxc.harvard.edu/proposer/POG/.}
in 2003 \citep{bam05a,pan10a} and 2008 (Table~\ref{tab01}).  The data for
these observations were reprocessed with version 4.6 of the CIAO suite of
analysis tools%
\footnote{More information about CIAO is available at
  http://cxc.harvard.edu/ciao/.}
and with version 4.6.1.1 of the CALDB%
\footnote{More information about the CALDB is available at
  http://cxc.harvard.edu/caldb/.}
to use the most recent data reduction algorithms and calibration products.
This process involved the following steps, in sequence.
The Level 1 event-data status bits 1--5 and 14--23 (of 0--31) were unset.
The tool {\tt destreak} was used to identify events associated with
horizontal streaks (electronic noise) on the ACIS-S4 CCD.
The tools {\tt acis\_build\_badpix} (first execution), {\tt
acis\_find\_afterglow}, and {\tt acis\_build\_badpix} (second execution)
were used to produce new observation-specific bad-pixel files.  Note that
the tool {\tt acis\_find\_afterglow} includes the most recent cosmic-ray
afterglow and hot-pixel identification algorithms.
The tool {\tt acis\_process\_events} was used to apply the new bad-pixel
files, to update the computation of the celestial coordinates, to update the
pulse-height information (i.e.\ the charge-transfer inefficiency (CTI) and
time-dependent gain adjustments), and to set certain status bits.
The tool {\tt dmcopy} was used to exclude events that are not in the
good-time intervals, that have a bad grade (1, 5, or 7), that have one or
more status bits set to one (e.g.\ occur on bad pixels or that are part of
an afterglow or horizontal ACIS-S4 streak), or that have an energy outside
the range from 1--5~keV.
The data at energies less than 1~keV were discarded because they are
dominated by emission from the foreground Vela supernova remnant
\citep{asc98a}.  The data at energies greater than or equal to 5~keV were
discarded because they are dominated by a charged-particle background.
There is no evidence of significant problems with the bias maps or of flares
in the particle background. The focal-plane temperature remained within
$1~^{\circ}{\rm C}$ of the nominal temperature of $-119.7~^{\circ}{\rm C}$.

\begin{figure}
  \includegraphics[width=6.5in]{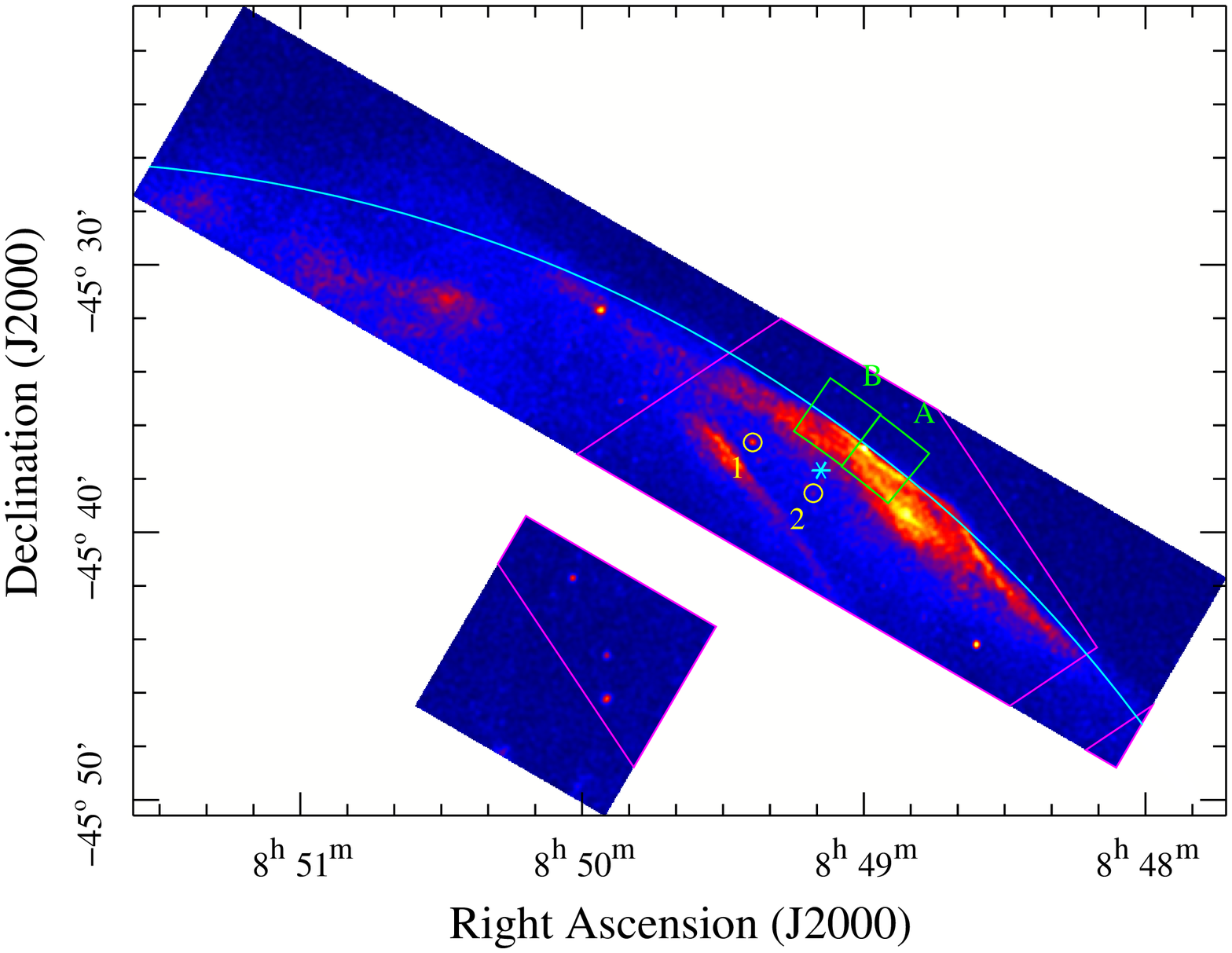}
  \caption{A 1--5~keV image of the northwestern rim of \snr\ from the 2003
    \cxo\ observation.
    The cyan asterisk is the location of the aim point.
    The image has been adjusted to compensate for instrumental effects, to
    the extent possible, and smoothed using a two-dimensional Gaussian
    function with $\sigma_{X} = \sigma_{Y} = 10~{\rm pixels} = 4.92$~arcsec.
    The color is a linear function of the flux and varies from about $1
    \times 10^{-9}$ or less (dark blue) to $1.4 \times 10^{-8}$ or more
    (white) in units of photons~cm$^{-2}$~s$^{-1}$~pixel$^{-1}$.
    The magenta lines mark the boundary of the region that was observed in
    both 2003 and 2008.
    The yellow circles encompass registration sources 1 and 2.
    The green annular wedges mark the boundaries of regions A and B, which
    were used to measure the rate of expansion.
    The cyan arc is a segment of a circle that has a radius of
    $0.8642^{\circ}$ and that is centered on the location of
    CXOU~J085201.4--461753 \citep{pav01a}.
    \label{fig01}}
\end{figure}

\begin{figure}
  \includegraphics[width=6.5in]{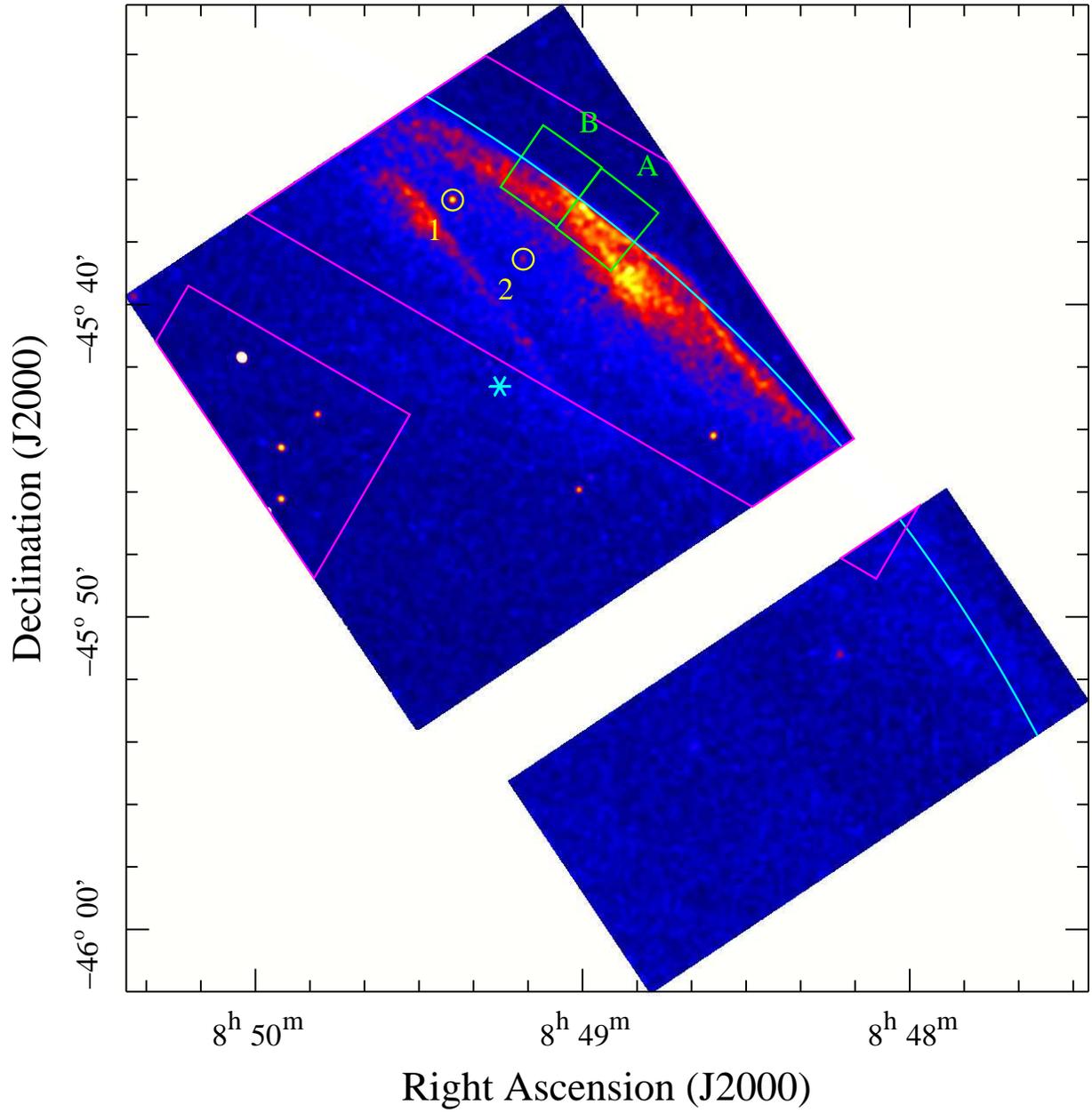}
  \caption{A 1--5~keV image of the northwestern rim of \snr\ from the 2008
    \cxo\ observation.  Refer to the caption of Figure~\ref{fig01} for more
    details.
    \label{fig02}}
\end{figure}

Figures~\ref{fig01} and \ref{fig02} are images of the 1--5~keV photon fluxes
incident on the \cxo\ telescope (i.e.\ the absorbed fluxes) in 2003 and
2008, respectively.  These images were produced using the following CIAO
tools in sequence: {\tt reproject\_events} (to place the data for OBS\_IDs
4414 and 9123 on the same celestial tangent plane as the data for OBS\_ID
3846), {\tt asphist}, {\tt mkinstmap} (with spectral weights from the
spectrum of the brightest portion of the region that was observed in both
epochs),%
\footnote{Spatial variations in the shape of the spectrum can lead to
  systematic errors in the relative flux.}
{\tt mkexpmap}, {\tt dmcopy}, and {\tt dmimgcalc}.  The point-spread
function of the \cxo\ mirrors and ACIS varies over the observed portion of
\snr.  At the aim point, 50\% of the events fall within a radius of about
$0.5\arcsec$. For a region that is $10\arcmin$ off axis, this radius is
about $5\arcsec$. Version 6.11 of the FTOOL%
\footnote{More information about FTOOLs is available at
  http://heasarc.nasa.gov/ftools/ftools\_menu.html.}
{\tt fgauss} was used to smooth the images with a two-dimensional Gaussian
function where $\sigma_{\rm X} = \sigma_{\rm Y} = 10\ {\rm pixels} =
4.92$~arcsec. This choice of Gaussian widths yields images that have
comparable spatial resolutions in the regions that were used to measure the
rate of expansion. While the procedure used to create Figures~1 and 2
removed most of the instrumental features, some residual artifacts remain,
particularly between CCDs and along the outer edges of the detectors.

\begin{figure}
  \includegraphics[width=5.5in]{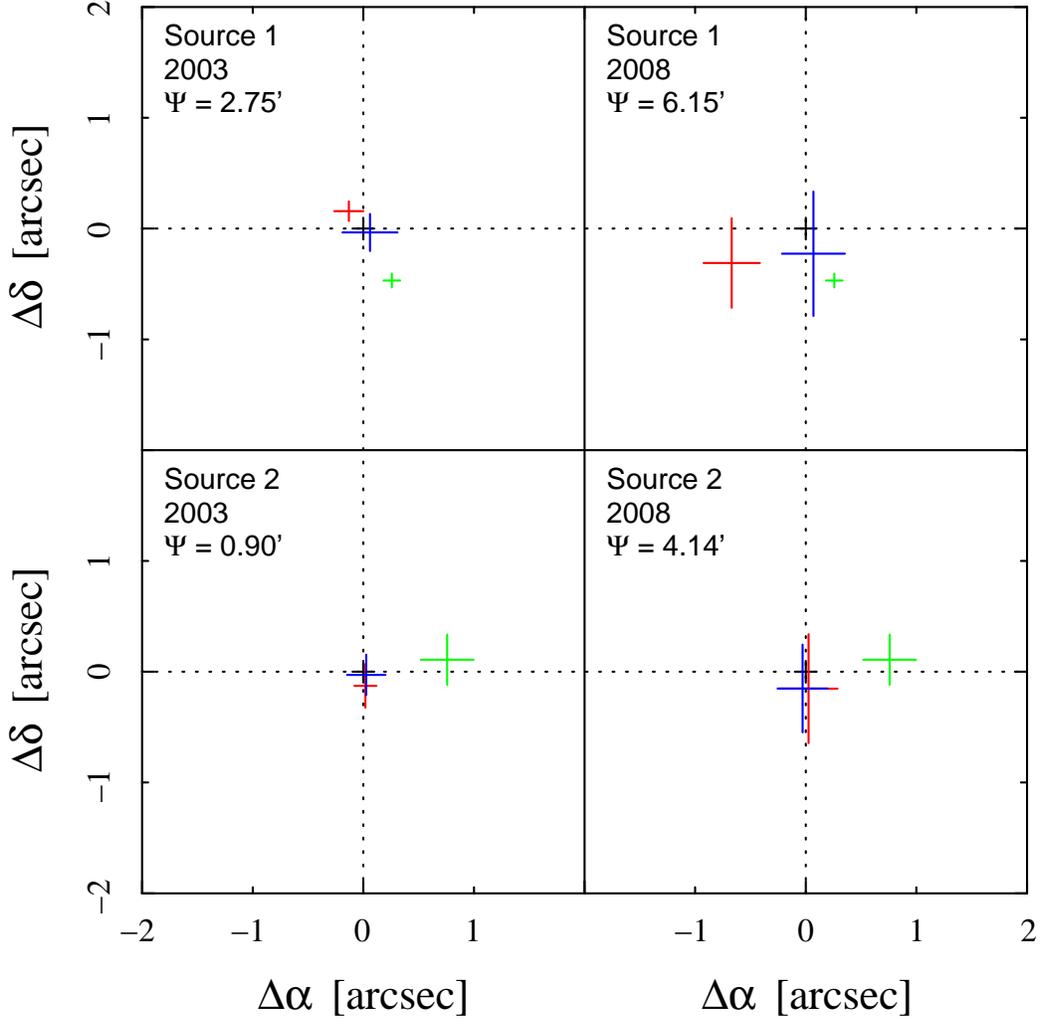}
  \caption{A comparison of the actual \cxo\ (red), expected \cxo\ (blue, see
    text), \twomass\ (black), and USNO-B1.0 (green) locations of the two
    registration sources. Here, $\Delta\alpha = \alpha_{\rm obs} -
    \alpha_{\twomass}$ and $\Delta\delta = \delta_{\rm obs} -
    \delta_{\twomass}$, where the subscript obs is either \cxo\ or
    USNO-B1.0.  The horizontal and vertical error bars denote the 90\%
    confidence level intervals.  The top and bottom panels are for source 1
    and source 2, respectively. The left and right panels are for the 2003
    and 2008 \cxo\ observations, respectively. The \twomass\ and USNO-B1.0
    locations do not change from the left side to the right side.  The
    angular separation between the registration source and the optical axis
    of \cxo\ is specified by $\Psi$ in arcminutes.
    \label{fig03}}
\end{figure}

The images were searched for potential registration sources. The two sources
listed in Table~\ref{tab02} and shown in Figures~\ref{fig01} and
\ref{fig02}, while faint, are point-like, are present in both the 2003 and
2008 datasets, seem to be spatially coincident with
sources in the \twomass%
\footnote{The \twomass\ All-Sky Point Source Catalog is available at
  http://www.ipac.caltech.edu/2mass/releases/allsky/.}
and USNO-B1.0%
\footnote{The USNO-B1.0 catalog \citep{mon03a} is available at
  http://www.usno.navy.mil/USNO/astrometry/optical-IR-prod/usno-b1.0.}
catalogs, and, at least for the 2003 observation, are moderately close to
the optical axis of \cxo.  These criteria give us some confidence that the
sources are not associated with diffuse emission from the remnant. The \cxo\
coordinates listed in Table~\ref{tab02} were computed for each source at
each epoch as follows.
A bin size in the range from 0.3 to 2.0 pixels (i.e.\ 0.15 to 0.98~arcsec)
was chosen.
The events located within a $60~{\rm bin} \times 60~{\rm bin}$ square grid
centered on the \twomass\ coordinates of the source were selected.
A one-dimensional, 60-bin histogram of these events was created along the
$X$ axis (i.e.\ Right Ascension) and fitted with a model that includes a
constant component for the background and a Gaussian component for the
source.  The four parameters for this model were allowed to vary.
A similar histogram was created along the $Y$ axis (i.e.\ Declination) and a
similar fit was performed.
This process was repeated for many bin sizes.  While the results at all bin
sizes are consistent with the results in Table~\ref{tab02}, in general the
results seem most reliable for bin sizes between about 0.5 and 1.0~pixels.
Therefore, the coordinates in this Table are the mean values of the results
obtained for bin sizes in this narrower range.
The \cxo\ and USNO-B1.0 coordinates are plotted relative to the \twomass\
coordinates in the four panels of Figure~\ref{fig03}. Since the \twomass\
coordinates are used for reference here, they are plotted at the centers of
the four panels.  The lengths of the horizontal and vertical bars for each
data point correspond to the 90\% confidence level intervals in Right
Ascension and Declination, respectively. The value of $\Psi$ in each panel
of the Figure denotes the angle between the optical axis of the \cxo\
telescope and the \twomass\ coordinates of the source. The wide range of
these angles (0.90 to 6.15 arcmin), suggests that the \cxo\ coordinates are
affected to varying degrees by asymmetries in the off-axis
point-spread function of the telescope.  {\sl MARX}%
\footnote{More information about the {\sl MARX} simulator is available at
  http://space.mit.edu/cxc/marx/.}
simulations were used to investigate this effect.  About $10^{7}$ simulated
X-rays were produced for each source at each epoch assuming that the sources
are at the locations given by the \twomass\ coordinates. A subset of the
simulated events were used to calculate the coordinates in the manner
described above. The number of events in each subset is consistent with the
number of events in the \cxo\ data.  Ten thousand subsets were processed for
each one of the four cases (i.e.\ each one of the four panels in
Fig.~\ref{fig03}).  The horizontal and vertical bars of the blue data points
in Figure~\ref{fig03} represent the intervals over which 90\% of the
simulated Right Ascension and Declination coordinates, respectively, are
found.  As shown, the simulated locations of the \twomass\ sources are
consistent with the actual \twomass\ locations.  Therefore, asymmetries in
the off-axis point-spread function of the \cxo\ telescope are not expected
to significantly affect the accuracy of the \cxo\ coordinates of the
registration sources. From Table~\ref{tab02}, the differences between the
\cxo\ coordinates of the registration sources in 2003 and the \twomass\
coordinates of the sources are $\Delta\alpha \equiv \alpha_{\cxo} -
\alpha_{\twomass} = -0.04 \pm 0.08$~arcsec and $\Delta\delta \equiv
\delta_{\cxo} - \delta_{\twomass} = 0.11 \pm 0.08$~arcsec. In 2008, the
differences are $\Delta\alpha = -0.33 \pm 0.18$~arcsec and $\Delta\delta =
-0.25 \pm 0.31$~arcsec. With the possible exception of source 1 in 2008,
(i.e.\ the upper, right-hand panel in Figure~\ref{fig03}), there is no
compelling evidence that the locations of the sources in the \cxo\ data are
inconsistent with the locations of the \twomass\ sources. For this reason,
the images shown in Figures~\ref{fig01} and \ref{fig02} have not been
adjusted to compensate for any potential registration errors.

\begin{figure}
  \includegraphics[width=6.5in]{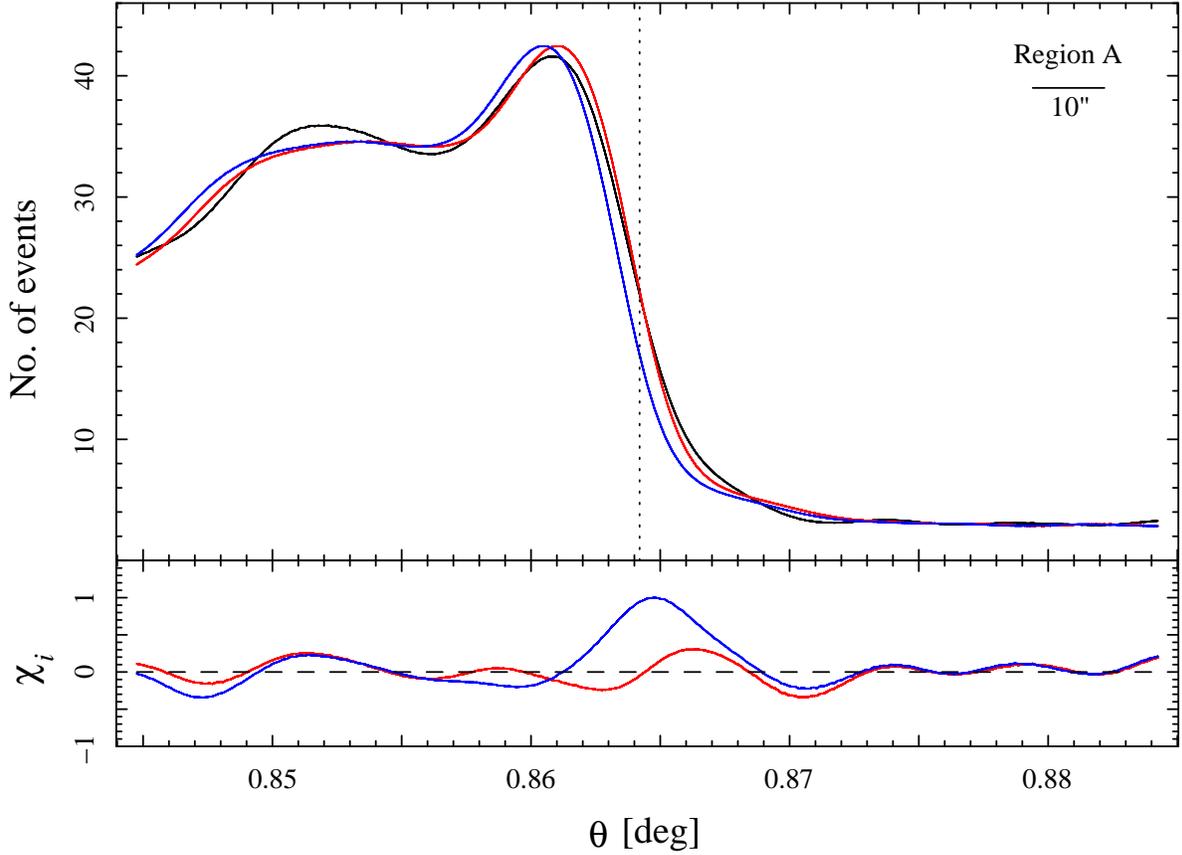}
  \caption{{\sl Top panel:} Radial profiles for region A (see
    Figures~\ref{fig01} and \ref{fig02}).  The black curve is the number of
    events in each radial bin from the 2008 dataset.  Here, the bins are 1
    pixel (0.492~arcsec) wide.  The dotted vertical line at $\theta =
    0.8642^{\circ}$ is the radius at which the number of events in 2008 is
    halfway between the peak of the black curve and the nominal number of
    background events at the right side of the profile.  The blue curve is
    the model before it has been radially shifted (i.e.\ $M$ from
    eqn.~\ref{eqn02} with $j=0$). A comparison of the blue and black curves
    shows that the change in the location of the forward shock from 2003 to
    2008 is evident.  The red curve is identical to the blue curve, except
    that it has been shifted to the right by 2.0~arcsec (Table~\ref{tab03}).
    For comparison, the line segment in the upper, right-hand corner is
    10~arcsec in length. {\sl Bottom panel:} The differences between the
    black profile and the blue and red profiles divided by the 1-$\sigma$
    statistical uncertainties (see eqn.~\ref{eqn03}).
    \label{fig04}}
\end{figure}

After determining that the 2003 and 2008 datasets are registered to an
accuracy of about 0.5~arcsec (1~pixel) or better, the rate of expansion was
measured in annular wedges A and B (Figures~\ref{fig01} and \ref{fig02}) as
follows.
For each epoch, an image was created in the sky coordinates $X$ and $Y$.
These images have 1~pixel $\times$ 1~pixel (i.e.\ 0.492~arcsec $\times$
0.492~arcsec) bins and were smoothed with a two-dimensional Gaussian
function where $\sigma_{X} = \sigma_{Y} = 10\ {\rm pixels} = 4.92$~arcsec.%
\footnote{Several values for the Gaussian width $\sigma = \sigma_{X} =
\sigma_{Y}$ were tried from 5 to 20~pixels.  While there is some indication
that the measured values of the radial offset $\Delta \theta$ decline with
increasing smoothing size $\sigma$, the best-fit values seem to be
insensitive to the smoothing size for $\sigma \le 10$~pixels.  Note that in
no case were the best-fit values inconsistent with the values listed in
Table~\ref{tab03}.  In fact, the fitted parameters were well within the 90\%
confidence intervals even at $\sigma = 20$~pixels.}
This choice of Gaussian widths yields 2003 and 2008 images that have
comparable spatial resolutions in regions A and B.
The two sky images were used to create images in the coordinates $\theta$
and $\phi$.  Here, $\theta$ is the angular radius (or angular separation) of
a point in the \cxo\ data from the assumed location of the center of the
remnant.  Since the location at which the progenitor exploded is unknown,
the center is assumed to be the location of the source
CXOU~J085201.4--461753 \citep[$\alpha_{\rm J2000} = 8^{\rm h}\, 52^{\rm m}\,
01\fs38$ and $\delta_{\rm J2000} = -46^{\circ}\, 17'\,
53\farcs34$,][]{pav01a}, which may be a compact object associated with \snr.
For calculations that depend upon the value of $\theta$, a 20\% uncertainty
is assumed because the remarkable correlation between the outer edge of the
shock and a circular arc with CXOU~J085201.4--461753 as the focus
(Figs.~\ref{fig01} and \ref{fig02}) becomes rather poor if the location the
focus (and the corresponding angular radius) is shifted by 20\% or more of
the angular radius. The angle $\phi$ is the azimuth. It is measured from
north ($\phi = 0^{\circ}$) through east ($\phi = 90^{\circ}$) relative to
the center of the remnant.  The images in $\theta$ and $\phi$ also have
0.492~arcsec $\times$ 0.492~arcsec bins.
Radial profiles were obtained from the $\theta$ and $\phi$ images by summing
along the $\phi$ direction.
The radial offset of the 2008 data with respect to the 2003 data was
obtained by interpolating each radial profile to a grid with a spacing of
$10^{-4}$~pixels (i.e.\ $4.92 \times 10^{-5}$~arcsec) and by minimizing the
fit statistic
\begin{equation}
  \chi^{2}
    =
    \sum_{i=1}^{n}
    \left(
      \frac{ C_{{\rm 2008},i} - M_{i} }{ \sigma_{i} }
    \right)^{2},
    \label{eqn01}
\end{equation}
where the ``model''
\begin{equation}
  M_{i}
    =
    s
    \left( C_{{\rm 2003},i-j} - \Sigma_{B,{\rm2003}} \Omega_{i-j} \right) +
    \Sigma_{B,{\rm 2008}} \Omega_{i},
    \label{eqn02}
\end{equation}
where the statistical uncertainty
\begin{equation}
  \sigma_{i}
    =
    \left( C_{{\rm 2008},i} + s^{2} C_{{\rm 2003},i-j} \right)^{1/2},
    \label{eqn03}
\end{equation}
and where $C_{{\rm 2003},i}$ and $C_{{\rm 2008},i}$ are the total number of
events in the $i$th bin of the interpolated radial profiles for the 2003 and
2008 data, respectively,
$\Sigma_{B,{\rm 2003}}$ and $\Sigma_{B,{\rm 2008}}$ are the number of
background events per square arcsec for the 2003 and 2008 data,
respectively,
$\Omega_{i}$ is the number of square arcseconds in the $i$th bin,
$s$ is a scaling factor, which compensates for differences in the detector
efficiencies, observing times, and source fluxes in the two epochs, and
$j$ is the radial offset of the 2008 data relative to the 2003 data in units
of $10^{-4}$~pixels.
In equation~\ref{eqn03}, the statistical uncertainty includes contributions
for both datasets.
During the fitting process, the value of $\chi^{2}$ was calculated using
many sets of values for the two variables $s$ and $j$, both of which were
allowed to vary freely.  The values of $\Sigma_{B,{\rm 2003}}$ and
$\Sigma_{B,{\rm 2008}}$ were frozen at the values listed in
Table~\ref{tab03}.%
\footnote{Some fits were performed with the values of $\Sigma_{B,{\rm
2003}}$ and $\Sigma_{B,{\rm 2008}}$ allowed to vary freely.  In these cases,
the best-fit values for these two parameters and for $s$ and $j$ were
consistent with the values listed in Table~\ref{tab03}.  Since the value of
$\chi^{2}$ seems to be rather insensitive to the values of $\Sigma_{B,{\rm
2003}}$ and $\Sigma_{B,{\rm 2008}}$ and since there was some difficulty
obtaining meaningful confidence level uncertainties for these two
parameters, their values were frozen in all other fits.}
These values were obtained from the numbers of events in and the sizes of
the source free portions of the two regions. The best-fit results for
regions A and B are listed in Table~\ref{tab03}. This Table includes the
parameter $\Delta \theta = 4.92 \times 10^{-5} j$. Since $j$ is quantized in
units of $10^{-4}$~pixels, $\Delta \theta$ is in units of arcsec.  The
uncertainties are listed at the 90\% confidence level, which corresponds to
a change in $\chi^{2}$ of 2.71.  Both of the parameters $s$ and $j$ were
allowed to vary while the confidence intervals were being computed.
Figure~\ref{fig04} shows profiles of $C_{\rm 2008}$, $M$ (with $\Delta
\theta = 0$~arcsec), and $M$ (with $\Delta \theta = 2.0$~arcsec) for region
A. Figure~\ref{fig05} shows the 1-, 2-, and 3-$\sigma$ confidence contours
(i.e.\ where $\chi^{2}$ changes by 2.30, 6.18, and 11.83) in the parameter
space defined by $\Delta \theta$ and $s$ for the same region.

There is significant evidence that the shock front of \snr\ expanded from
2003 to 2008, at least in regions A and B.  The amounts of expansion for
these two regions are consistent with one another at the 90\% confidence
level (Table~\ref{tab03}) and are insensitive to the mean registration
adjustments (e.g.\ Figure~\ref{fig05}).  Since there is no compelling
evidence that the registration of the \cxo\ data is inaccurate, the results
presented hereafter are the results obtained without registration
adjustments and the uncertainties are quoted at the 90\% confidence level.
The mean amount of expansion for regions A and B is $\Delta \theta = 2.40
\pm 0.56$~arcsec over a period of 5.652~yr, which corresponds to an
expansion rate of $\dot{\theta} = 0.42 \pm 0.10$~arcsec~yr$^{-1}$.  If the
shock radius $\theta = 0.86^{\circ}$, the angular distance between
CXOU~J085201.4--461753 and the northwestern rim (Fig.~\ref{fig04}), then the
fractional expansion rate $\dot{\theta} / \theta = 0.136 \pm
0.042$~kyr$^{-1}$.

If the flux from regions A and B of \snr\ did not change, then the scaling
factors $s$ should be consistent with the ratios $\Sigma_{B,{\rm 2008}} /
\Sigma_{B,{\rm 2003}}$ of the background event densities.  These ratios are
$0.331 \pm 0.048$ and $0.332 \pm 0.049$, respectively, for regions A and B,
which are consistent with the best-fit values of $s$.  Therefore, there is
no evidence of significant flux changes from 2003 to 2008 for these two
regions.


\section{Discussion}
\label{dis}

\subsection{Age}
\label{age}

The expansion rate measured using \cxo\ data ($0.42 \pm
0.10$~arcsec~yr$^{-1}$) is about half the expansion rate obtained using {\sl
XMM-Newton} data \citep[$0.84 \pm 0.23$~arcsec~yr$^{-1}$,][]{kat08a}.  The
two measurements are from similar time intervals (2003 to 2008 for \cxo\ and
2001 to 2007 for {\sl XMM-Newton}) and from similar regions (i.e.\ bright
portions of the northwestern rim).  However, the regions used are not
identically the same.  It is possible that there is an azimuthal variation
in the expansion rate along the northwestern rim of \snr.  Although the
difference is not statistically significant, the expansion rates in regions
A and B differ by a factor of 1.5 (Table~\ref{tab03}).  While there may not
be significant differences in the expansion rates along the northeastern rim
of SN~1006 \citep{kat09a}, there do appear to be significant azimuthal
variations in the expansion rates of Cas~A \citep{del03a} and Kepler
\citep{kat08b}.  Furthermore, the variations reported for these latter two
remnants could be as large as a factor of two or more.

If the angular radius $\theta = 0.86^{\circ}$, then the fractional expansion
rates $\dot{\theta}/\theta$ are $0.136 \pm 0.042$~kyr$^{-1}$ and $0.27 \pm
0.07$~kyr$^{-1}$ for \cxo\ and {\sl XMM-Newton}, respectively.%
\footnote{\cite{kat08a} assume an angular radius of $1^{\circ}$ instead of
  $0.86^{\circ}$.  Therefore, they report a fractional expansion rate of
  $0.23 \pm 0.06$~kyr$^{-1}$ instead of the rate of $0.27 \pm
  0.07$~kyr$^{-1}$ that we use for comparison.}
The fractional expansion rate provides a crude constraint on the age of the
remnant. If the radius of the forward shock $r_{\rm f} \propto t^{m}$, where
$t$ is the age and $m$ is the expansion parameter, and if $m$ is a constant
over the time interval of the expansion measurement, then $t = m \theta /
\dot{\theta}$.  While the value of $m$ is unknown, it is most likely in the
range from 0.4 (the Sedov-Taylor phase) to 1 (the
free-expansion phase). In this case, the age of \snr\ is in the range%
\footnote{These ranges include the 90\% confidence level uncertainty on
  $\dot{\theta}/\theta$.}
from 2.1 to 13~kyr (\cxo) or from 1.2 to 5.0~kyr ({\sl XMM-Newton}),
provided the expansion results for the northwestern rim are representative
of the remnant as a whole.

To try to obtain better constraints on the age of \snr, we used the
hydrodynamic models of \cite{tru99a}. Since the physical conditions---the
initial kinetic energy ($E_{0}$), mass ($M_{\rm ej}$), and mass density
distribution ($\rho_{\rm ej} \propto v^{-n} t^{-3}$) of the ejecta, the
ambient mass density distribution ($\rho_{0} = 1.42 m_{\rm p} n_{0}$), and
the evolutionary state ($t/t_{\rm ch}$)---are unknown, a five-dimensional
grid in $E_{0}$, $M_{\rm ej}$, $n$, $n_{0}$, and $t/t_{\rm ch}$ was used
with 81 values of $E_{0}$: $10^{49}$, $10^{49.05}$, {\ldots},
$10^{53}$~ergs; 41 values of $M_{\rm ej}$: $10^{0}$, $10^{0.05}$, {\ldots},
$10^{2}$~$M_{\odot}$; seven values of $n$: 6, 7, 8, 9, 10, 12, 14 (i.e.\ $m
= (n-3)/n = 0.50$, {\ldots}, 0.79); 101 values of $n_{0}$: $10^{-5}$,
$10^{-4.95}$, {\ldots}, $10^{0}$~cm$^{-3}$; and 999 values of $t/t_{\rm
ch}$: 0.01, 0.02, {\ldots}, 9.99.  Note that the characteristic age $t_{\rm
ch} = 6.18\, (E_{0} / 10^{51}~{\rm ergs})^{-1/2} (M_{\rm ej} / 10
M_{\odot})^{5/6} (n_{0} / 0.1~{\rm cm}^{-3})^{-1/3}$~kyr. Collectively, the
grid includes 2.35 billion scenarios.  Of course, most of the scenarios are
improbable.  We tried to make the range of values for each parameter large
enough to bracket the expected value of the parameter.  The following four
criteria were used to determine which scenarios and, hence, which ages, are
plausible.

One criteria is that the distance-independent fractional expansion rate
(i.e.\ $v_{\rm f} / r_{\rm f} = \dot{\theta} / \theta$) must be compatible
with the \cxo\ result (i.e.\ is in the range from 0.094 to
0.178~kyr$^{-1}$).  Scenarios that did not satisfy this criteria were
discarded.

Another criteria is that the forward shock speed must be greater than or
equal to 1000~km~s$^{-1}$.  Such speeds are required \citep[e.g.\ eqn.~A5
of][]{all08a} to accelerate electrons to energies high enough to produce
X-ray synchrotron spectra with cut-off frequencies in excess of $10^{17}$~Hz
\citep{pan10a}.

A third criteria is that the inferred amount of thermal X-ray emission from
the forward-shocked material cannot exceed observational constraints. The
pshock model of XSPEC was used to describe this emission. The abundances
were assumed to be solar and the temperature was set to 0.3~keV. While the
temperature is unknown, it is expected to be at least this high since the
electron temperatures measured for other X-ray synchrotron-emitting remnants
are larger. Had a higher temperature been used, then more scenarios would
have been discarded. The lower and upper limits on the pshock ionization
timescale were set to zero and $1.2 n_{0} t$, respectively.  The pshock
normalization was set to $3.64 \times 10^{-18} (n_{0}/1\ {\rm cm}^{-3})^{2}
(r_{\rm f}/1\ {\rm cm})$~cm$^{-5}$.  This value is based upon the
assumptions that the shock-heated gas fills one quarter of the volume inside
the forward shock and that it has an electron-to-proton ratio of 1.2 (i.e.\
that the proton and electron densities are $4 n_{0}$ and $4.8 n_{0}$,
respectively). The pshock emission is absorbed using the XSPEC model tbabs
with $n_{\rm H} = 1.2 \times 10^{22}$~atoms~cm$^{-2}$. This absorption
column density represents an upper limit for the remnant \citep[][]{ace13a}.
Lower column densities would have resulted in more scenarios being
discarded. The absorbed thermal X-ray spectrum was compared to the total
{\sl ROSAT} spectrum for a 0.86$^{\circ}$ radius cone along the line of
sight through \snr. This spectrum includes emission from both the \snr\ and
Vela supernova remnants. As long as the absorbed emission model for a
scenario does not exceed the {\sl ROSAT} spectrum by more than $3 \sigma$ at
any point in the spectrum, the scenario is considered plausible. This
constraint limits the ambient density to be below 0.4~cm$^{-3}$.

The last criteria used is an energy constraint. The sum of the kinetic and
thermal energies of the forward-shocked material plus the inferred energy of
the cosmic-ray protons cannot exceed $E_{0}$, the initial kinetic energy of
the ejecta. This constraint is less restrictive than it would have been if
the computation also included the energies associated with the shocked and
unshocked ejecta, with the other cosmic-ray particles, and with the magnetic
field. The kinetic energy of the forward-shocked material is assumed to be
given by $U_{\rm KE,f} = 3 \pi \rho_{0} r_{\rm f}^{3} v_{\rm f}^{2} / 8$
(i.e.\ $= M_{\rm s} (3 v_{\rm f} / 4)^{2} / 2$, where $M_{\rm s} = 4 \pi
\rho_{0} r_{\rm f}^{3} / 3$). The thermal energy of the forward-shocked
material is assumed to be given by the same expression (i.e.\ $U_{kT,{\rm
f}} = \sum_{i} 3 N_{i} kT_{i} / 2$, where $N_{i} = 4 \pi \rho_{i} r_{\rm
f}^{3} / ( 3 m_{i} )$ and $kT_{i} = 3 m_{i} v_{\rm f}^{2} / 16$).  The total
energy in cosmic-ray protons is obtained by integrating the
momentum-dependent energy over the power-law number-density distribution
$dn/dp = A (p/p_{0})^{-\Gamma} \exp((p_{0} - p)/p_{\rm max})$ and by
multiplying by one quarter of the total volume.  Here, $A = 3.5 \times
10^{-8}$~cm$^{-3}$~(GeV$/c$)$^{-1}$, $p_{0} = 1$~GeV$/c$, $\Gamma = 2.0$,
and $p_{\rm max} = 18$~TeV$/c$.  These parameters are based on a joint fit
of an inverse Compton model to gamma-ray data and of a synchrotron model to
radio and X-ray data \citep[see][]{all11a}.  The number density of protons
is assumed to be one hundred times larger than the number density of
electrons at $p = 1$~GeV$/c$.  A larger proton-to-electron ratio would have
lead to more scenarios being discarded. The integration is performed from
$p_{\min}$ to $10 p_{\rm max}$.  The former quantity is the momentum at
which there is a transition from a Maxwell-Boltzmann distribution to the
power-law. Implicit in this calculation is the requirement that the number
density of thermal protons must exceed the number density of nonthermal
protons at thermal momenta.  The energy constraint limits $E_{0}$ to be
above $4 \times 10^{49}$~ergs, $n_{0}$ to be above $3 \times
10^{-4}$~cm$^{-3}$, $r_{\rm f}$ to be less than 70~pc (i.e.\ $d < 5$~kpc),
and $v_{\rm f}$ to be less than $10^{4}$~km~s$^{-1}$.

Of the 2.35 billion scenarios considered, 57.4 million (2.45\%) satisfy all
four of our plausibility criteria.  A histogram of the ages for the
plausible scenarios is shown in Figure~\ref{fig06}.  The youngest plausible
scenario has an age of 2.2~kyr.  The oldest is 8.4~kyr. If the lowest 5\%
and highest 5\% of the distribution are discarded, then the 90\% confidence
level interval for the age is from 2.4 to 5.1~kyr.

This range is based upon the assumption that the models of \cite{tru99a} are
suitable for \snr. Although \cite{tru99a} only considered uniform ambient
densities, it is possible to obtain a simple scaling factor between the age
obtained using their model and the age expected if the ambient material has
a mass density distribution $\rho \propto r^{-s}$. At early times (i.e.\
those for which $t < 0.5\, t_{\rm ch}$), the effective value of $m$ would be
given by $(n-3)/(n-s)$, not $(n-3)/n$ \citep[][]{tru99a}. As a result, $m$,
and hence the age, is larger by a factor of $n/(n-s)$. Therefore, if the
ambient material is from a steady wind (i.e.\ $s = 2$) and if $n \ge 6$,
then the age is underestimated by no more than a factor of 1.5.

The remnant was most likely produced by a core collapse supernova:%
\footnote{\cite{iyu05a} argue that \snr\ was produced by a sub-Chandrasekhar
  Type Ia supernova.  However, this argument is based, in part, on the
  unlikely assumption that the remnant emits an observable flux of 1.16~MeV
  gamma rays associated with the decay of $^{44}$Ti \citep{iyu98a}.}
(1) There are reports of an X-ray emitting compact central object
\citep{asc98a,sla01a,pav01a}. (2) There is no evidence of thermal X-ray
emission, which suggests that \snr\ is expanding into the rarefied
environment of a stellar wind-blown bubble \citep{sla01a,lee13a}. (3) There
is a molecular cloud (the Vela Molecular Ridge) and a group of massive stars
\citep[i.e.\ Vel~OB1,][]{egg82a} with which the remnant may be associated.
Yet, if \snr\ is the remnant of a Type Ia supernova, instead of a core
collapse event, then \cite{dwa98a} suggest the ejected material may have an
exponential mass density distribution $\rho_{\rm ej} \propto e^{-v} t^{-3}$
instead of a power-law distribution $\rho_{\rm ej} \propto v^{-n} t^{-3}$.
Although \cite{tru99a} did not consider models with exponential ejecta
profiles, \cite{dwa98a} note that Type Ia remnants have been modeled using a
power law with $n = 7$.  If the sample of plausible scenarios is limited to
the subset with $n = 7$ and with $M_{\rm ej} = 1.4 M_{\odot}$, then the 90\%
confidence level interval for the age is from 2.4 to 4.5~kyr with no
scenario having an age less than 2.2~kyr or more than 6.1~kyr.

For these reasons, the age of \snr\ is expected to be between 2.4 and
5.1~kyr whether or not it was produced by a core collapse supernova. In no
case is the remnant expected to be younger than 2.2~kyr, which contradicts
most of the previous age estimates. These estimates are reviewed hereafter.

The first estimate published was that of \cite{asc98a}, who argues that
\snr\ is less than or about 1.5~kyr old.  This result is based upon the high
temperature that is obtained when the {\sl ROSAT} PSPC data are fitted with
a thermal emission model. However, subsequent observations with the {\sl
ASCA} GIS \citep{tsu00a}, which had better spectral resolution, reveal that
the X-ray flux is dominated by synchrotron radiation and show no evidence of
thermal emission \citep{sla01a}.

Several age estimates are based upon evidence of emission associated with
the decay of $^{44}$Ti. \cite{iyu98a} report an emission line at 1.16~MeV in
{\sl COMPTEL} data and obtain an age of about 0.68~kyr. \cite{che99a} and
\cite{asc99a} expand upon this work and find ages between 0.6 and 1.1~kyr
and less than 1.1~kyr, respectively.  \cite{tsu00a} report the detection of
a $4.1 \pm 0.2$~keV X-ray emission line from the northwestern rim with {\sl
ASCA}.  They attribute this line to $^{44}$Ca produced by the decay of
$^{44}$Ti. Based upon the 1.16~MeV line flux, they estimate an age between
0.6 and 1.0~kyr.  \cite{sla01a} reexamined the {\sl ASCA} data. While they
find a hint of a 4~keV line in the SIS0 data for a region in the northwest,
they find no such evidence in the SIS1 data for the same region.
\cite{iyu05a} report a line feature at $4.45 \pm 0.05$~keV in {\sl
XMM-Newton} spectra for the northwestern, western, and southern rims.
\cite{hir09a} find no evidence of a 4~keV emission line in {\sl Suzaku}
data.  Their upper limits on the line flux are well below the line fluxes
reported by \cite{tsu00a}, \cite{iyu05a}, and \cite{bam05a}.  Their limits
are also below the X-ray line flux inferred from the gamma-ray line flux of
\cite{iyu98a}. Furthermore, there is some concern about the statistical
significance of the 1.16~MeV line in the {\sl COMPTEL} data \citep{sch00a}
and there is no evidence of 67.9 and 78.4~keV lines in the {\sl INTEGRAL}
data \citep{ren06a}.  Since the evidence of X-ray and gamma-ray emission
lines associated with the decay of $^{44}$Ti is questionable, claims of
their detections do not provide a compelling reason to doubt the age range
inferred from the measurement of the expansion rate.

Following suggestions that \snr\ is young, \cite{bur00a} searched for
evidence of a geophysical signature of the supernova that produced it. They
report that South Pole ice core samples exhibit temporal spikes in the
abundance of nitrate and that these spikes may be associated with historic
supernovae.  Their Figure~1 shows spikes that could be associated with the
Kepler, Tycho, and AD~1181 supernovae.  It also shows a spike that occurred
in AD $1320 \pm 20$.  If this spike is associated with \snr, then the age of
the spike would be consistent with ages inferred from the reports of
$^{44}$Ti line emission.  However, they note that it is not clear that the
ionizing radiation from supernovae significantly affect the terrestrial
nitrate abundance. For example, there is no spike associated with Cas~A.
Unfortunately, the results presented in their figure do not go back far
enough to determine whether or not there are spikes associated with the Crab
and SN~1006 supernovae.

\cite{obe14a} performed a hydrodynamic analysis assuming that \snr\ is
expanding into an environment with several molecular clouds that have a
variety of masses and densities.  While they note that they cannot eliminate
the possibility that the remnant is a few thousand years old, they favor an
age of about 0.8~kyr.  Unfortunately, it's not clear that their assumptions
are applicable to \snr.  For example, there's no clear evidence that the
remnant is expanding into a medium with a large density gradient. In their
simulated images, the X-ray emission from the rim of the remnant is
irregularly shaped and clumpy, unlike the observations, which show an outer
edge that follows a nearly circular arc in the northwest (Figs.~\ref{fig01}
and \ref{fig02}).  Furthermore, the X-ray emission in their model is
entirely thermal, whereas the X-ray emission from the remnant is dominated
by nonthermal emission, at least above about 1~keV.  In fact, the conditions
of their model (i.e.\ a downstream shock-heated plasma with $n =
1$~cm$^{-3}$ and $kT = 1$~keV) are incompatible with the limits obtained
from the {\sl ROSAT} data. If $kT = 1$~keV, then the downstream density
cannot be larger than 0.4~cm$^{-3}$.  For these reasons, the results of
their analysis may not be accurate for \snr.

\cite{bam05a} infer an age between 0.42 and 1.4~kyr using a novel technique.
This technique is based upon a simple hydrodynamic model and upon a
relationship between the age and the quantity $\nu_{\rm roll} / l^{2}$,
where $\nu_{\rm roll}$ and $l$ are the roll-off frequency and downstream
scale length of X-ray synchrotron radiation, respectively. At present, there
is too little data to evaluate its reliability.

Of the previously-published ages, 17.5~kyr \citep{tel09a} is unique in that
it is larger than the ages inferred from the \cxo\ data. Telezhinsky
suggests that \snr\ entered the radiative phase 1.5~kyr ago.  As a result, a
significant fraction of the thermal energy of the shock-heated gas has been
lost.  Therefore, it's possible to have a large ambient density ($n_{0} =
1.5$~cm$^{-3}$) without the thermal X-ray emission being higher than
observational constraints.  The large density means that the TeV gamma-ray
emission can be described primarily in terms of the photons produced by
neutral-pion decay instead of inverse Compton scattering.  However, it's not
clear that this model is consistent with the detection of thin,
X-ray--synchrotron-dominated filaments. These filaments, which are thought
to be associated with the forward shock, have synchrotron cut-off
frequencies in excess of $10^{17}$~Hz \citep{pan10a}.  Using Telezhinsky's
magnetic field strength of 67~$\mu$G, the corresponding electron cut-off
energy is greater than 10~TeV. Yet, if particle acceleration at the forward
shock ended 1.5~kyr ago \citep{tel09a}, then electrons with energies in
excess of 1~TeV would have lost their energy via synchrotron radiation. The
synchrotron spectrum would have a cut-off frequency of about $10^{15}$~Hz
and would not detectable at X-ray energies.  Another concern is that the
model may violate energy conservation. For example, the amount of energy
transferred to the bulk kinetic motion and to the random thermal motion of
the forward-shocked material (even if some of this energy has been lost via
radiation) may be expressed as $U_{\rm f} \equiv U_{\rm KE,f} + U_{kT,{\rm
f}} = 3 \pi \rho_{0} r_{\rm f}^{3} v_{\rm f}^{2} / 4 = 1.96 \times 10^{53}
(n_{0}/1~{\rm cm}^{-3}) (\theta / 1^{\circ})^{3} (\dot{\theta} / 1~{\rm
arcsec}~{\rm yr}^{-1})^{2} (d / 1~{\rm kpc})^{5}~{\rm ergs}$.  Using
Telezhinsky's values of $n_{0}$ (1.5~cm$^{-3}$) and $d$ (0.6~kpc) and our
values of $\theta$ (0.86$^{\circ}$) and $\dot{\theta}$
(0.42~arcsec~yr$^{-1}$), yields $U_{\rm f} = 2.57 \times 10^{51}$~ergs.
Although this computation excludes all other forms of energy, the value of
$U_{\rm f}$ is still much larger than Telezhinsky's initial kinetic energy
of $2 \times 10^{50}$~ergs.

In summary, a wide range of ages have been inferred for \snr\ using a
variety of different evidence.  Of these reports, we argue that the most
reliable are those based upon the measurement of the expansion rate of the
remnant. The biggest concern with this technique is that the expansion rate
in the northwest may not be representative of the remnant as a whole. If the
\cxo\ results are accurate and representative, then \snr\ is between 2.4 and
5.1~kyr old.

\subsection{Distance}

The measured expansion rate, even when coupled with the hydrodynamic
simulations, does not provide a significant constraint on the distance.  For
example, Figure~\ref{fig07} shows that the range from 0.7 to 3.2~kpc
encompasses 90\% of the 57.4~million scenarios that satisfy the plausibility
criteria described in \S\ref{age}, The full range of distances for these
scenarios is from 0.3 to 4.5~kpc. Hereafter, we review other inferences
about the distance.

If, as expected, the remnant was produced by a core collapse supernova, then
it is likely that it is part of the Vela Molecular Ridge. This material is
concentrated into two groups \citep{mur91a}, one at a distance of $0.7 \pm
0.2$~kpc \citep{lis92a} and the other at a distance of about 2~kpc. In
particular, the progenitor of \snr\ may have been a member of either the
Vel~OB1 or Vel~OB2%
\footnote{The Vel~OB2 association of \cite{egg82a} should not be confused
  with an entirely different association referred to as Vela~OB2 by
  \cite{dez99a}.}
associations, which are at distances of about 0.8 and 1.8~kpc, respectively
\citep{egg82a}. \cite{dun00a} report that a distance of 1--2~kpc, instead of
a distance much less than 1~kpc, yields a diameter that is more compatible
with the diameters of remnants that have similar radio surface brightnesses
(see their Fig.~6).

\cite{rey06a} use the column density $n_{\rm H}$ toward the central compact
object CXOU~J085201.4--461753, as measured by \cite{bec02a}, to infer a
distance of $2.4 \pm 0.4$~kpc.  However, \cite{ace13a} show that most of the
column density along this line of sight (and toward the remnant) is
associated with the Vela Molecular Ridge. Therefore, the compact object (and
\snr) can be no further than 0.9~kpc. This limit is consistent with the
results of \cite{kim12a}, who suggest that the remnant may be a source of
far-ultraviolet emission, in which case it is closer than 1~kpc.

Since the values of $n_{\rm H}$ associated with \snr\ are significantly
larger than the values associated the Vela supernova remnant \citep{sla01a},
the remnant must lie beyond Vela \citep[$d_{\rm Vela} = 0.29 \pm
0.02$~kpc,][]{dod03a}. A more restrictive lower limit can be obtained from
the properties of the X-ray synchrotron emission in the northwest. Since the
synchrotron cut-off frequency exceeds $10^{17}$~Hz \citep{pan10a}, the shock
speed must be larger than about 1000~km~s$^{-1}$ \citep{all08a}, which
implies that the distance $d = v_{\rm f} / \dot{\theta} > 0.5$~kpc.

Conversely, \cite{asc13a} argues that the remnant is closer, perhaps much
closer, than 0.5~kpc.  This argument hinges upon the assumption that the
gamma-ray emission is dominated by neutral-pion decay.%
\footnote{Although we use the same energy constraints as \cite{asc13a}, we
assume that the TeV gamma rays are dominated by inverse Compton scattering
instead of neutral-pion decay.  In this case, it is possible for the remnant
to be considerably more distant than 0.5~kpc.}
Yet, \cite{ber09a} could not make such a model work if the remnant is
nearby.  Furthermore, if the {\sl Fermi} spectrum of \cite{tan11a} includes
emission from both \snr\ and PSR~J0855$-$4644, particularly at the lower end
of their spectrum, then an inverse Compton scattering model may provide a
better description of the gamma-ray emission from \snr\
\citep[e.g.][]{kat08a,lee13a}.

Early estimates of the distance, based upon evidence of line emission
associated with the decay of $^{44}$Ti, also suggest the source is nearby
[e.g.\ $d = 0.2$~kpc, \citep{iyu98a}; $d = 0.1$--0.3~kpc, \citep{che99a};
and $d < 0.5$~kpc, \citep{asc99a}]. However, as described in \S\ref{age},
this evidence is questionable.

In summary, the preponderance of the distance results suggests that \snr\ is
between about 0.5 and 1~kpc from Earth.  Hereafter, we assume the remnant is
at a distance of $0.7 \pm 0.2$~kpc.  Note that this assumption does not
significantly affect the constraints on the age. Table~\ref{tab04} lists
sample hydrodynamic properties for \snr, assuming it is at a distance of
either 0.5, 0.7, or 0.9~kpc. These properties include the initial kinetic
energy, $E_{0}$, the mass, $M_{\rm ej}$, and the power-law index, $n$, of
the ejected material, the ambient density, $n_{0}$, the age, $t$, the
radius, $r_{\rm f}$, and speed, $v_{\rm f},$ of the forward shock, the
expansion parameter, $m$, the mass of the material swept up by the forward
shock, $M_{\rm s}$, and the amount of kinetic energy that has been
transferred to this material, $U_{\rm KE,f}$. The values in the Table, which
are based upon the results of the hydrodynamic study described in
\S\ref{age}, are only meant to be representative.  The values for each
property can vary substantially from the listed values for individual
scenarios.  The value of $n$ was arbitrarily chosen to be nine.  As a
result, the value of $m = 2/3$. The value of $E_{0}$ was arbitrarily chosen
to be $10^{51}$~ergs, except for the case with $d = 0.5$~kpc. At the closer
distance, which is at the low end of the distribution in Figure~\ref{fig07},
none of the scenarios with $E_{0} = 10^{51}$~ergs (or with $E_{0} >
10^{51}$~ergs) satisfied all of the plausibility criteria.


\section{Conclusions}
\label{concl}

\begin{figure}
  \includegraphics[width=6.5in]{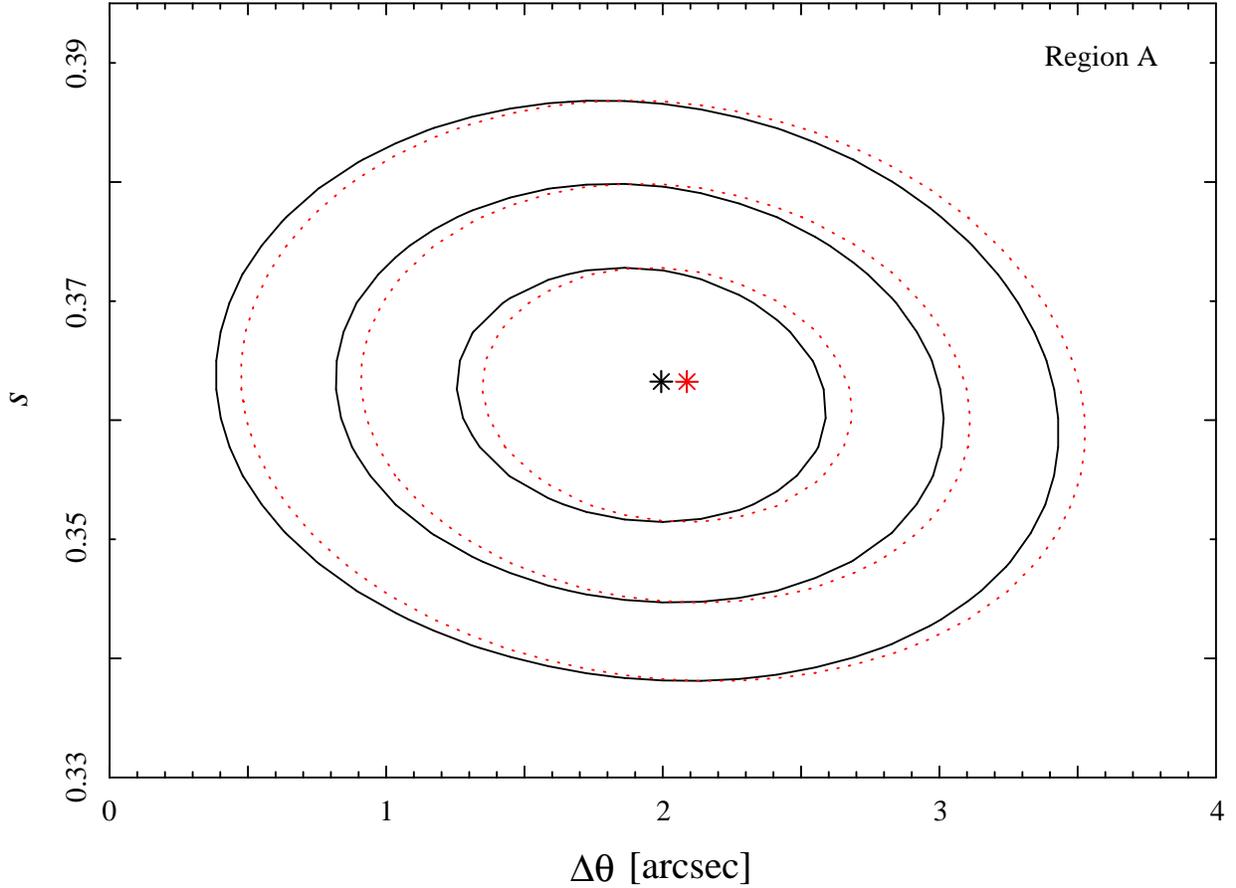}
  \caption{The 1-, 2-, and 3-$\sigma$ confidence contours for region~A (see
    Figures~\ref{fig01} and \ref{fig02}) in the parameter space defined by
    the radial expansion $\Delta \theta$ and the scaling factor $s$.  The
    solid black and dotted red contours are the results obtained before and
    after, respectively, the mean $\Delta\alpha$ and $\Delta\delta$
    registration adjustments (see sec.~\ref{data}) are included. The stars
    indicate the best-fit values of $\Delta \theta$ and $s$.  The evidence
    of expansion in region A (i.e.\ the evidence that $\Delta\theta > 0$) is
    significant at nearly the 4-$\sigma$ confidence level.
    \label{fig05}}
\end{figure}

\begin{figure}
  \includegraphics[width=6.5in]{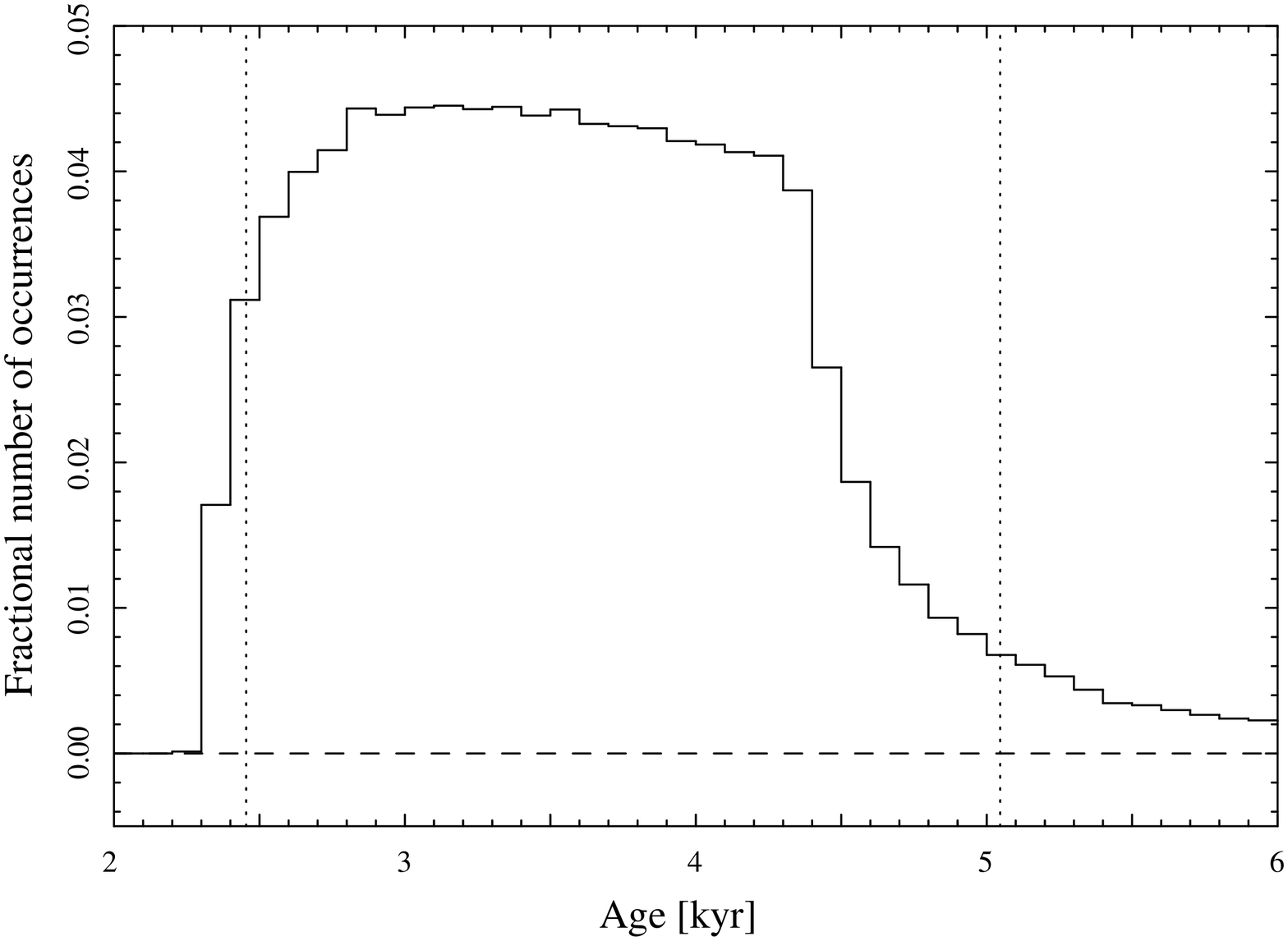}
  \caption{The distribution of the ages of the 57.4 million plausible
    hydrodynamic scenarios described in \S\ref{age}.  If the lowest 5\% and
    highest 5\% of the distribution are ignored, then the plausible ages lie
    between about 2.4 and 5.1 kyr (i.e.\ between the dotted vertical lines).
    \label{fig06}}
\end{figure}

\begin{figure}
  \includegraphics[width=6.5in]{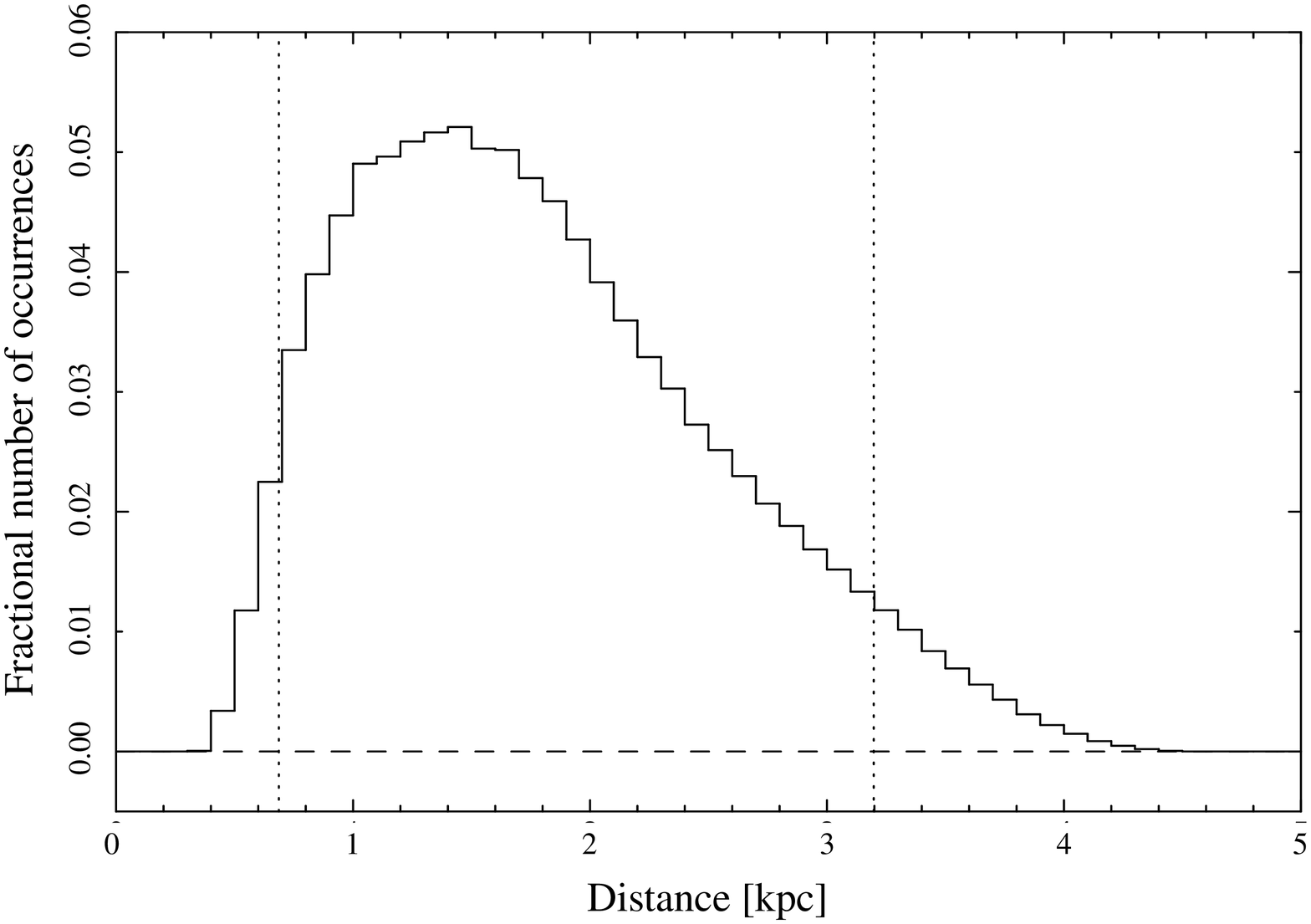}
  \caption{The distribution of the distances of the 57.4 million plausible
    hydrodynamic scenarios described in \S\ref{age}.  If the lowest 5\% and
    highest 5\% of the distribution are ignored, then the plausible
    distances lie between about 0.7 and 3.2 kpc (i.e.\ between the dotted
    vertical lines).
    \label{fig07}}
\end{figure}

We reprocessed and analyzed our 2003 and 2008 \cxo\ ACIS data for the
supernova remnant \snr\ to search for evidence of expansion
\citep[e.g.][]{kat08a}.  Two objects satisfy our criteria for potential
registration sources. Instrumental simulations reveal no evidence of a
significant registration error.  For this reason, and because the expansion
results are insensitive to small registration errors (Fig.~\ref{fig05}), no
registration adjustments were applied.  The data for two adjacent annular
wedges along a relatively bright and narrow portion of the northwestern rim
indicate that it has experienced a radial displacement of about $2.40 \pm
0.56$~arcsec over a period of 5.652 yr.  The corresponding expansion rate
($0.42 \pm 0.10$~arcsec~yr$^{-1}$ or $13.6 \pm 4.2$\%~kyr$^{-1}$) is about
half of the rate reported for an analysis of {\sl XMM-Newton} data from a
similar time interval and a similar region \citep{kat08a}. Since the regions
used are not identical, one possible explanation for this difference is an
azimuthal variation in the expansion rate. Additional observations would
provide a more precise measure of the mean expansion rate and enable a
search for azimuthal variations.

To constrain the age, a hydrodynamic analysis was performed using the models
of \cite{tru99a}. Billions of scenarios were considered using broad ranges
of initial kinetic energies, ejecta masses, ejecta mass density
distributions, ambient densities, and evolutionary states to try to
encompass all possible sets of hydrodynamic properties. Of these scenarios,
57.4~million are considered plausible because their properties are
consistent with the \cxo\ expansion rate (assuming it is representative of
the remnant as a whole), an inferred lower limit on the forward shock speed
(1000~km~s$^{-1}$), an inferred upper limit on the thermal X-ray emission,
and an energy constraint.  Ninety percent of the plausible scenarios have
ages in the range from 2.4 to 5.1~kyr.  The age of \snr\ is most likely in
this range whether or not it was produced by a Type Ia or Type II event. If
the remnant is expanding into the material shed by a steady stellar wind
instead of a uniform ambient medium, then it could be older by a factor of
up to 1.5. In no case is the remnant expected to be younger than 2.2~kyr.
Since the measurements of the expansion rate seem to provide a more reliable
means of determining the age than other techniques that have been used (see
\S\ref{age}), \snr\ is most likely too old to be associated with emission
from the decay of $^{44}$Ti or with features in the abundance of nitrate in
South Pole ice core samples.

We set a lower limit on the distance of 0.5~kpc.  This limit is based upon
the \cxo\ expansion rate and the requirement that the shock speed be greater
than or equal to 1000~km~s$^{-1}$.  (The detection of X-ray synchrotron
emission is not expected for lower shock speeds.) An analysis of
previously-published distance estimates and constraints suggests that the
remnant is no more than 1.0~kpc from Earth. Therefore the distance of \snr\
is consistent with the distance of the closer of two groups of material in
the Vela Molecular Ridge \citep[i.e.\ $0.7 \pm 0.2$~kpc,][]{lis92a}. This
distance is also consistent with the progenitor having been a member of the
Vel~OB1 association \citep{egg82a} and with our estimates of the age range.
Note that constraining the distance does not significantly affect the
estimate of the age.


\acknowledgments
We thank the anonymous referee, whose comments helped improve the
manuscript.
G.E.A.\ is supported by contract SV3-73016 between MIT and the Smithsonian
Astrophysical Observatory.
The Smithsonian Astrophysical Observatory is operated on behalf of NASA
under contract NAS8-03060.
This research has made use of data products from the Chandra Data Archive,
the Two Micron All Sky Survey, and the USNO-B1.0 catalog.
The analyses described herein were performed using the software package
CIAO, provided by the Chandra X-ray Center, the software package ISIS
\citep{hou00a}, the scripting language S-Lang, and models of the XSPEC
spectral-fitting package.


{\it Facilities:} \facility{CXO (ACIS)}.



\clearpage

\clearpage

\clearpage

\clearpage

\clearpage

\clearpage

\clearpage

\clearpage

\begin{deluxetable}{lcc}
\tablecaption{\cxo\ observations of the northwestern rim\label{tab01}}
\tablewidth{0pt}

\tablehead{
  \colhead{\ } &
  \colhead{2003} &
  \colhead{2008}
}

\startdata
Observation IDs
  & 3846, 4414
  & 9123
  \\
Start date
  & 2003 Jan 5
  & 2008 Aug 31
  \\
End date
  & 2003 Jan 7
  & 2008 Sep 1
  \\
Duration [ks]
  & 74
  & 40
  \\
Pointing location:
  &
  &
  \\
\hspace*{0.25in} RA [J2000]
  & 8$^{\rm h}$ 49$^{\rm m}$ 9.40$^{\rm s}$
  & 8$^{\rm h}$ 49$^{\rm m}$ 15.34$^{\rm s}$
  \\
\hspace*{0.25in} Dec [J2000]
  & -45$^{\circ}$ 37$'$ 42.4$''$
  & -45$^{\circ}$ 42$'$ 37.7$''$
  \\
ACIS detectors used\tablenotemark{a}
  & I2, S0, S1, S2, S3, S4
  & I0, I1, I2, I3, S2, S3
  \\
Maximum effective area\tablenotemark{b} [cm$^{2}$]
  & 700 @ 1.5~keV
  & 575 @ 1.5~keV
  \\
Effective energy band\tablenotemark{c} [keV]
  & 0.4--7.5
  & 0.6--7.8
  \\
Fractional energy resolution\tablenotemark{d} [FWHM/$E$]:
  &
  &
  \\
\hspace*{0.25in} At 1~keV
  & 0.10
  & 0.13
  \\
\hspace*{0.25in} At 5~keV
  & 0.03
  & 0.05
  \\
\enddata

\tablenotetext{a}{Each $1024\ {\rm pixel} \times 1024\ {\rm pixel}$ ACIS CCD
  has a field of view of 8.4~arcmin $\times$ 8.4~arcmin.}

\tablenotetext{b}{The effective area is a function of energy and position.
  The values reported here are the largest values at the locations of the
  aim points.  The maximum effective area declines away from these
  locations.}

\tablenotetext{c}{Here, the effective energy band is the range over which
  the effective area at the aim point is greater than or equal to 10\% of
  the maximum effective area.}

\tablenotetext{d}{The fractional energy resolution is a function of energy
  and position.  The values reported here are the values at the locations of
  the aim points.}

\end{deluxetable}


\clearpage

\begin{deluxetable}{lcccccc}
\rotate
\tablecaption{Registration sources\label{tab02}}
\tablewidth{0pt}

\tablehead{
  \colhead{\ } &
  \multicolumn{3}{c}{Source 1} &
  \multicolumn{3}{c}{Source 2}
  \\
  \cline{2-4}
  \cline{5-7}
  \colhead{\ } &
  \colhead{\ } &
  \colhead{$\alpha$\tablenotemark{a}} &
  \colhead{$\delta$\tablenotemark{a}} &
  \colhead{\ } &
  \colhead{$\alpha$\tablenotemark{a}} &
  \colhead{$\delta$\tablenotemark{a}}
  \\
  \ &
  \colhead{ID} &
  \colhead{[h m s]} &
  \colhead{[$^{\circ}$ $'$ $''$]} &
  \colhead{ID} &
  \colhead{[h m s]} &
  \colhead{[$^{\circ}$ $'$ $''$]}
}

\startdata
\cxo\ 2003 &
  1 &
  08 49 23.935(13) &
  $-$45 36 39.59(9) &
  2 &
  08 49 11.041(19) &
  $-$45 38 33.51(20)
  \\
\cxo\ 2008 &
  1 &
  08 49 23.883(24) &
  $-$45 36 40.06(40) &
  2 &
  08 49 11.041(47) &
  $-$45 38 33.53(49)
  \\
\twomass\tablenotemark{b} &
  08492394-4536397 &
  08 49 23.947(16) &
  $-$45 36 39.75(10) &
  08491103-4538333 &
  08 49 11.039(16) &
  $-$45 38 33.38(10)
  \\
USNO-B1.0\tablenotemark{c} &
  0443-0146834 &
  08 49 23.971(7) &
  $-$45 36 40.22(6) &
  0443-0146664 &
  08 49 11.111(23) &
  $-$45 38 33.27(23)
  \\
\enddata

\tablenotetext{a}{The coordinates are in the J2000 epoch.  The numbers in
  parentheses are the 90\% confidence level uncertainties in units of
  $10^{-3}$ seconds of Right Ascension or $10^{-2}$ arcseconds of
  Declination.}

\tablenotetext{b}{This information is from the \twomass\ All-Sky Point Source
  catalog.$^{9}$}

\tablenotetext{c}{This information is from the USNO-B1.0 catalog.$^{10}$}

\end{deluxetable}


\clearpage

\begin{deluxetable}{lcc}
\tablecaption{Expansion results\label{tab03}}
\tablewidth{0pt}

\tablehead{
  \colhead{Quantity\tablenotemark{a}} &
  \colhead{Region A} &
  \colhead{Region B}
}

\startdata
Region boundaries: & & \\
  \hspace*{0.25in} $\theta$ [deg] & 0.8442--0.8842 & 0.8447--0.8847 \\
  \hspace*{0.25in} $\phi$ [deg] & 320.0--322.5 & 322.5--325.0 \\
  & & \\
Model parameters: & & \\
  \hspace*{0.25in} $\Delta\theta$ [arcsec] & $1.98 \pm 0.72$ &
    $3.03 \pm 0.89$ \\
  \hspace*{0.25in} $s$ & $0.363 \pm 0.011$ & $0.324 \pm 0.012$ \\
  \hspace*{0.25in} $\Sigma_{B,{\rm 2003}}$ [events arcsec$^{-2}$] &
    $0.138 \pm 0.003$\tablenotemark{b} & $0.127 \pm 0.003$\tablenotemark{b} \\
  \hspace*{0.25in} $\Sigma_{B,{\rm 2008}}$ [events arcsec$^{-2}$] &
    $0.046 \pm 0.002$\tablenotemark{b} & $0.042 \pm 0.002$\tablenotemark{b} \\
  & & \\
Expansion: & & \\
  \hspace*{0.25in} $\Delta t$ [yr] & 5.652 & 5.652 \\
  \hspace*{0.25in} $\dot{\theta} = \Delta\theta / \Delta t$ [arcsec yr$^{-1}$]
    & $0.35 \pm 0.13$ & $0.54 \pm 0.16$ \\
  \hspace*{0.25in} $\theta$ [deg] & $0.86 \pm 0.17$\tablenotemark{c} &
    $0.86 \pm 0.17$\tablenotemark{c} \\
  \hspace*{0.25in} $\dot{\theta} / \theta$ [kyr$^{-1}$] &
    $0.113 \pm 0.047$ & $0.172 \pm 0.061$ \\
\enddata

\tablenotetext{a}{The statistical uncertainties are listed at the 90\%
  confidence level.}

\tablenotetext{b}{These values are based upon the sizes of and numbers of
  events in the source free portions of the regions.}

\tablenotetext{c}{A 20\% uncertainty in the shock radius is assumed because
  the location at which the progenitor exploded is unknown.}

\end{deluxetable}


\clearpage

\begin{deluxetable}{lccc}
\tablecaption{Sample hydrodynamic properties\label{tab04}}
\tablewidth{0pt}

\tablehead{
  \colhead{} &
  \multicolumn{3}{c}{Assumed distance} \\
  \cline{2-4}
  \colhead{Property} &
  0.5 kpc &
  0.7 kpc &
  0.9 kpc
}

\startdata
$E_{0}$ [$10^{51}$~ergs] &
  0.5 &
  1.0 &
  1.0 \\
$M_{\rm ej}$ [$M_{\odot}$] &
  50 &
  40 &
  28 \\
$n$ &
  9 &
  9 &
  9 \\
$n_{0}$ [cm$^{-3}$] &
  0.022 &
  0.022 &
  0.018 \\
$t$ [kyr] &
  3.9 &
  4.2 &
  5.2 \\
$r_{\rm f}$ [pc] &
  8 &
  11 &
  14 \\
$v_{\rm f}$ [$10^{3}$~km~s$^{-1}$] &
  1.3 &
  1.7 &
  1.7 \\
$m$ &
  0.67 &
  0.67 &
  0.67 \\
$M_{\rm s}$ [$M_{\odot}$] &
  1.4 &
  3.9 &
  6.6 \\
$U_{\rm KE,f}$ [$10^{50}$~ergs] &
  0.23 &
  0.60 &
  1.1 \\
\enddata

\end{deluxetable}

\end{document}